\documentclass[12pt]{article}
\newcommand{\bt}{\begin{theorem}}
\newcommand{\ds}{\displaystyle}
\newcommand{\et}{\end{theorem}}
\newcommand{\bea}{\begin{eqnarray}}
\newcommand{\eea}{\end{eqnarray}}

\def \spec#1 {\mathop{#1}}

\def \b #1 {\bf #1}
\newcommand {\be}{\begin{equation}}
\newcommand {\ee}{\end{equation}}

\newcommand{\vsp}{\vskip 1em}

\newcommand{\hsp}{\hskip 2em}

\newcommand{\ben}{\begin{eqnarray*}}
\newcommand{\een}{\end{eqnarray*}}

\def \qed {\hfill \vrule height6pt width6pt depth0pt}
\newtheorem{lemma}{Lemma}
\newtheorem{theorem}[lemma]{Theorem}

\def \noi{\noindent}

\def \geq{\ge}
\def \leq{\le}
\newcommand{\bdsc}{\begin{description}}
\newcommand{\edsc}{\end{description}}

\def\ba{\begin{array}}
\def\ea{\end{array}}

\title{Cooperative oligopoly games: a probabilistic approach}

\begin{document}
\date{}
\author{Paraskevas Lekeas{\footnote{Department of Applied Mathematics, University of Crete, 71409 Heraklion, Crete, Greece; email:
plekeas@tem.uoc.gr\vspace{0.2cm}}}\hsp Giorgos
Stamatopoulos{\footnote{Department of Economics, University of
Crete, 74100 Rethymno, Crete, Greece; email:
gstamato@econ.soc.uoc.gr\vspace{0.2cm}}}}

\maketitle

\begin{abstract}\noi We analyze the core of a cooperative Cournot game.
We assume that when contemplating a deviation, the members of a coalition assign
positive probability over all possible coalition structures
that the non-members can form. We show that when the number of firms in the market is sufficiently large then
the core of the underlying cooperative game is non-empty. Moreover, we show that the core of our game 
is a subset of the $\gamma$ core.
\end{abstract}

\vspace{0.2cm} \noi\hspace{0.83cm} {\small{\emph{Keywords}:
Cooperative game with externalities; Cournot market; core

\vspace{0.05cm}\hspace{0.3cm} bounded rationality


\section{Introduction}
The issue of cooperation among firms in oligopolistic markets constantly attracts the interest of economists.
Among other avenues, economists analyze cooperation in the market by examining the non-emptiness of
the core of an appropriately defined cooperative game. The core consists of all these allocations of
total market profits that cannot be blocked by any coalition of firms. When the members of a coalition contemplate to block (or
not) an allocation they need to calculate the worth of their coalition. In a market environment such a calculation is
not a trivial task, though, as the coalition's worth depends on how the non-members would act. Namely, it depends on
the coalition structure that the outsiders will form.

Different beliefs about the reaction of the outsiders lead to different notions of core.
The $\alpha$ and $\beta$ cores (Aumann 1959) are based on min-max behavior on behalf of the non-members;
the $\gamma$ core (Chander and Tulkens 1997) is based on the assumption that outsiders play individual best replies to the
deviant coalition.  Various authors applied these core notions to the study of oligopolistic markets. Rajan (1989)
used the concept of $\gamma$ core and showed that it is non-empty
for a market with 4 firms. Currarini and Marini (1998) built a refinement of the $\gamma$ core by assuming that
the deviant coalition acts as a Stackelberg leader in the product market. Zhao (1999) showed that the $\alpha$ and $\beta$
cores of oligopolistic markets are non-empty.

The seminal works of Bloch (1996) and Ray $\&$ Vohra (1999) go one step further, as in their approaches
the reactions of the outsiders --and the resulting coalition structures-- are deduced via equilibrium arguments.
However, Sandholm et.al (1999) showed that for an $n$-player game the number of different coalition structures
is $O(n^n)$ and $\omega(n^{\frac{n}{2}})$. Hence
computing which coalition structure the outsiders form is, in general, a particularly difficult task
(in fact the problem is $NP$-complete).

The last result gives the motivation  of the current paper. We analyze the core of a Cournot oligopoly
assuming that no group of firms has the computational ability to accurately deduce the
coalition structure that other firms will form. Instead, when a coalition contemplates a deviation from the
grand coalition it assumes that all possible partitions of the outsiders can arise with
positive probability. As a first step, we assume that coalition structures are all equiprobable.
By imposing a uniform distribution over outsiders' reactions, the current paper can be seen as offering
a boundedly rational approach to the literature of cooperative games with externalities.

We derive the worth function of any coalition using the above scenario and we examine the core of
the corresponding game. We focus in a market with linear demand and cost functions and product homogeneity.
Our main result says that when the number of firms is sufficiently large
then no coalition has incentive to break full cooperation; hence the core is non-empty.
We also examine the relation of our core with the $\gamma$ core and we show that the former
core is a subset of the latter.

In what follows, we present the model in section 2 and in section 3 we present our results.
Section 4 provides concluding remarks.

\section{Model}
We consider a market with the set $N=\{1,2,...,n\}$ of firms. Firms produce a homogeneous product facing the inverse demand function
$P=max\{a-Q,0\}$ where $P$ is the market price,
$Q=q_1+q_2+...+q_n$ is the market quantity, $q_l$ is the quantity of firm $l$, $l=1,2,...,n$ and $a>0.$
Firm $l$ produces with the cost function $C(q_l)=cq_l, l=1,2,...,n,$ where $c<a$.

Let $S\subseteq N$ denote a coalition with $|S|=s$ firms and let $N/S$ denote the complementary set of $S$.
The value (worth) of $S$ is the sum of its members' profits. In order to compute this value, the members of $S$ need to
predict how the members of $N/S$ partition themselves into coalitions. The set $N/S$ can partition into
disjoint subsets in $B_{n-s}$ possible ways, where $B_{n-s}$ is Bell's $(n-s)^{th}$
number (Bell (1934)). The $B_{n-s}$ different partitions define $B_{n-s}$
different coalition structures that coalition $S$ might face in the market. In this paper we incorporate the assumption of bounded computational abilities of agents with regard to which structure will form. One way to do this is by assuming that the members of coalition $S$ treat all structures as equiprobable, with probability $\frac{1}{B_{n-s}}$.

Let $j$ denote a coalition structure with $j$ members (coalitions), $j=1,2,...,n-s$. Observe that all
coalition structures with $j$ members induce the same profit for $S$ (as firms are symmetric).
The number of coalition structures with $j$ members is
$k_j$, where $k_j$ is given by the coefficients
of the Bell polynomials $B_{n-s,j}$ (the Stirling numbers of the second kind)
which give the number of ways to partition a set of $n-s$ objects into $j$ groups:

\be\label{Stirling second kind} k_j={n-s \brace j}=\frac{1}{j!}\sum\limits_{i=0}^j (-1)^i {{j}\choose{i}}(j-i)^{n-s} \ee

\noi What matters for $S$ is the number of coalitions it faces in the market (if two coalition structures
have the same number of firms they induce the same profit for $S$). So define the function
\be\label{ProbabilityDistribution} f_{n,s}(j)=\frac{k_j}{B_{n-s}},\hspace{0.2cm}j=1,2,...,n-s \ee
\noi which is the probability that a coalition $S$ with $s$ members assigns to
coalition structures with $j$ coalitions, $j=1,2,...,n-s$. With a slight abuse of notation, let $q_i^j$ denote the
quantity that coalition $i$ chooses, $i=1,2,...,j,$ under structure $j$. Let also $q_s$ denote the quantity
of coalition $S$. The profit function that coalition $S$ faces is then given by

\be\label{prof S} \pi(S)=\sum\limits_{j=1}^{n-s}f_{n,s}(j)(a-q_s-\sum\limits_{i=1}^j q_i^j-c)q_s\ee

\noi Moreover, from the perspective of coalition $S$, the profit function of coalition $i$ within structure $j$ is
$$\label{one} \pi_i^j=(a-q_s-\sum\limits_{i=1}^j q_i^j-c)q_i^j,\hspace{0.2cm}i=1,2,...,j,\hspace{0.1cm}j=1,2,...,n-s$$

\noi Hence the maximization problems to
solve for are

$$\label{max for S} max_{q_s}\pi(S)$$

$$\label{max for ij} max_{q_i^j}\pi_i^j,\hspace{0.2cm}i=1,2,...,j,\hspace{0.1cm}j=1,2,...,n-s$$

\noi By symmetry, the solution of the above problems will involve $q_1^j=q_2^j=...=q_j^j\equiv q^j$, $j=1,2,...,n-s$. Define
\be\label{F} {\ds{F_{f_{n,s}}=\sum\limits_{j=0}^{n-s}\frac{j\cdot f_{n,s}(j)}{j+1}}} \ee
\noi Then it is easy to show that \be\label{qsfinal} q_s=\frac{1-F_{f_{n,s}}}{2-F_{f_{n,s}}}(a-c)\ee

\noi and for $j=1,2,...,n-s$,

\be\label{qjfinal} q_i^j=q^j=\frac{a-c}{(j+1)(2-F_{f_{n,s}})},\hspace{0.2cm}\noi i=1,2,...,j\ee

\noi Using (\ref{qsfinal}),(\ref{qjfinal}) and (\ref{Stirling second kind}) in (\ref{prof S}), we obtain $v(S)$ as{\footnote{Normally, we should
sum from $j=1$ up to $n-s$ but for convenience we start from $j=0$ with the understanding that $f_{n,s}(0)=0$.}} 



\be\label{ff worth S} v(S)=\frac{(\alpha-c)^2}{B_{n-s}}\frac{1-F_{f_{n,s}}}{(2-F_{f_{n,s}})^2}\sum \limits_{j=0}^{n-s} \frac{{n-s \brace j}}{j+1}\ee

\noi Hence our game is the pair $(N,v)$ where $v$ is defined by (\ref{ff worth S}).

\section{Properties of the game}

\noi In this section we derive some important properties of the game $(N,v)$.

\noi \begin{lemma}{\label{population}} Let $v^n(s)$ denote the value for coalition $S \neq \emptyset$
with $|S|=s$ in a game with $n$ players. For every positive integer $k$ we have that $v^n(s)=v^{n+k}(s+k)$. \end{lemma}

\vspace{0.2cm}\noi \textbf{Proof.} $v^n(s)=\frac{(\alpha-c)^2}{B_{n-s}}\cdot\frac{1-F_{f_{n,s}}}{(2-F_{f_{n,s}})^2}\cdot\sum
\limits_{j=0}^{n-s}\frac{{n-s \brace j}}{j+1}=$

~\

$=\frac{(\alpha-c)^2}{B_{(n+k)-(s+k)}}\cdot\frac{1-F_{f_{n,s}}}{(2-F_{f_{n,s}})^2}\cdot \sum \limits_{j=0}^{(n+k)-(s+k)} \frac{{(n+k)-(s+k) \brace j}}{j+1}=v^{n+k}(s+k)$

~\

\noi because from (\ref{F}) $F_{f_{n+k,s+k}}=F_{f_{n,s}}$. \qed

~\

\noi An almost immediate implication of Lemma \ref{population} is the monotonicity of $v(S)$ in $|S|=s$.

\begin{lemma}{\label{strictly_monotone}} For every $S$ with $|S|=s\leq n$, $v^n(S)$ is strictly increasing in $s$.\end{lemma}

\noi \textbf{Proof.} We will use induction on the number of players $n$. For the base case, $n=2$, we have to prove that $v^2(2)>v^2(1)>v^2(0)$.
We have that $v^2(2)=\left(\frac{a-c}{2}\right)^2>\left(\frac{a-c}{3}\right)^2=v^2(1)>0=v^2(0)$, so we have the base case. Assume for the induction hypothesis that in a game with $n$ players and for an arbitrary $s$, $1 < s \leq n$ we have that $v^n(s)>v^n(s-1)$. We will prove that $v^{n+1}(s)>v^{n+1}(s-1)$. But this is an immediate result of lemma \ref{population} and the induction hypothesis since $v^{n+1}(s)=v^n(s-1)>v^n(s-2)=v^{n+1}(s-1)$. \qed

~\

\noi We are now ready to state and prove the main result of this section.

\vsp\noi \textbf{Proposition 1.} \emph{The game $(N,v)$ has non-empty core for all $n \geq 11$.}

\vspace{0.2cm}\noi \textbf{Proof.} Since firms are identical, we can use Lemma 1 of Rajan (1989), according
to which the core of a game is non empty if and only if for all $S:|S|=s \leq n$

\be\label{9-Rajan}\frac{v(n)}{n} \geq \frac{v(s)}{s}\ee

\noi It is easy to verify that inequality (\ref{9-Rajan}) does not hold for $3 \leq n \leq 10$,
so for these values of $n$ the core is empty.\footnote{For $3 \leq n \leq 10$ and for all
$S$ with $|S|=1$ it holds that $v^n(1)>\frac{v^n(n)}{n}$. See Table \ref{n=11_table} in the Appendix.} We will prove the rest of the proposition using induction on $n$, $n\geq 11$.

~\

\noi \textit{Base}: Table \ref{nonempty} in the Appendix establishes the base case ($n=11$).

~\

\noi \textit{Induction hypothesis}: For all $S:|S|=s \leq n$, $\frac{v^n(n)}{n} \geq \frac{v^n(s)}{s}$.

~\

\noi \textit{Induction step}: We will show that for all $S:|S|=s \leq n+1$, $$\frac{v^{n+1}(n+1)}{n+1} \geq \frac{v^{n+1}(s)}{s}$$

\noi By Lemma \ref{population} we have that $v^{n+1}(s)=v^{n+1}((s-1)+1)=v^n(s-1)$ and also that $v^{n+1}(n+1)=v^n(n)$. So we have to show that 

\be\label{inductionequation1} \frac{v^n(n)}{n+1} \geq \frac{v^n(s-1)}{s} \ee

\noi From the Induction hypothesis we have $$\label{inductionequation2} v^n(n) \geq \frac{n}{s-1} v^n(s-1) $$

\noi and thus \be\label{i1} (s-1)v^n(n)\geq nv^n(s-1)\ee

\noi Using Lemma \ref{strictly_monotone}, \be\label{i2}v^n(n)>v^n(s-1)\ee

\noi Adding (\ref{i1}) and (\ref{i2}) gives $ s v^n(n)\geq (n+1) v^n(s-1) $ which implies that (\ref{inductionequation1}) holds. So we have the proof for $n+1$ and thus the proposition. \qed

~\

\noi Our result shows that for low values of $n$, the sign of $v^n(n)/n-v^n(S)/s$ can be negative for some $s$, whereas for large $n$  the sign is always positive. To see why we need large $n$ for core non-emptiness, we note that as $n$ increases, both $v^n(n)/n$ and $v^n(S)/s$ decrease. However, the second term decreases faster than the first. Hence there would be an $n$ above which the difference $v^n(n)/n-v^n(S)/s$ eventually becomes positive for all $s$.

\subsection{Relation with the $\gamma$ core}
Let us now analyze the relation of our core with the $\gamma$ core. Under the latter notion,
the members of a deviant coalition believe that the outsiders play individual best replies.
Below we show that the $\gamma$ core is non-empty for our market and that our core is a subset of the $\gamma$
core. Let $v_{\gamma}(S)$ denote the worth function of $S$ under the scenario of $\gamma$-behavior
of players. The pair $(N,v_{\gamma})$ will denote the game under the latter scenario; finally
$C_{\gamma}$ will denote the core of $(N,v_{\gamma})$ and $C_f$ the core of our game.

\vsp\noi {\bf{Remark 1}} \emph{$C_{\gamma}\neq\emptyset$}

\vspace{0.2cm}\noi {\bf{Proof.}} It is straightforward to show that
${\ds{v_{\gamma}(S)=\frac{(a-c)^2}{(2+n-s)^2}}}$ and
that ${\ds{v_{\gamma}(N)=\frac{(a-c)^2}{4}}}$. Hence ${\ds{\frac{v_{\gamma}(N)}{n}\geq
\frac{v_\gamma(S)}{s}}}$ if and only if $sn^2+(4s-4-2s^2)n+s(4+s^2-4s)\geq 0$ which holds.
Hence the $\gamma$ core is non-empty.
\qed

\noi \begin{lemma} $C_f\subset C_{\gamma}$\end{lemma}

\noi \textbf{\textbf{Proof.}} By Proposition 1, if $n\in\{3,4, \cdots ,10\}$,
we have $C_f=\emptyset\subset C_\gamma$. If $n\geq 11$,
then $C_f\neq \emptyset.$ In this case, it suffices to show that $v(S)>v_\gamma(S)$.
To this end, let us give a useful representation of $v(S)$. The representation is based on
harmonic numbers. Recall that the $k$-th harmonic number is defined as $$\label{harmonic} h^k=1+\frac{1}{2}
+\frac{1}{3}+ \cdots +\frac{1}{k}=\sum\limits_{j=0}^{k-1}\frac{1}{1+j}$$ Let us now define
$$ \label{weighted harm}h_f^k=\sum\limits_{j=0}^{k-1}\frac{f(j)}{1+j}$$ as the \emph{$k$-th probabilistic harmonic number} where $f(.)$ is a probability distribution on $\{0,1,2, \cdots ,k\}$. To make a connection between the above concept and our game, notice that we can write (\ref{ff worth S}) as 

\be\label{oneharmonic v}v(S)=(a-c)^2 \frac{1-F_{f_{n,s}}}{(2-F_{f_{n,s}})^2}h_{f_{n,s}}^{n-s+1}\ee

\noi where $$h_{f_{n,s}}^{n-s+1}=\sum\limits_{j=0}^{n-s}\frac{f_{n,s}(j)}{1+j}$$

\noi Notice next that $$ \label{h1}{\ds{F_{f_{n,s}}=\sum\limits_{j=0}^{n-s}\frac{j \cdot f_{n,s}(j)}{j+1}=
\sum\limits_{j=0}^{n-s}[1-\frac{1}{j+1}]f_{n,s}(j)=1-\sum\limits_{j=0}^{n-s}\frac{f_{n,s}(j)}{j+1}}}$$

\noi Hence \be \label{h2}F_{f_{n,s}}=1-h_{f_{n,s}}^{n-s+1}\ee

\noi Then combining (\ref{oneharmonic v}) and (\ref{h2}) we get

\be\label{harmonic v}v(S)=\frac{(h_{f_{n,s}}^{n-s+1})^2}{(1+h_{f_{n,s}}^{n-s+1})^2}(a-c)^2\ee

\noi A similar representation can be given for $v_\gamma(S)$ as well. Let $h_{\gamma}^{n-s+1}$
denote the probabilistic harmonic number associated with $(N,v_\gamma)$. Then
$${\ds{h_\gamma^{n-s+1}=\frac{1}{n-s+1}}}$$
\noi and thus $$v_\gamma(S)=\frac{(h_\gamma^{n-s+1})^2}{(1+h_\gamma^{n-s+1})^2}(a-c)^2$$

\noi Hence in order to show that $v(S)>v_\gamma(S)$ it suffices to
show that $h_f^{n-s+1}>h_\gamma^{n-s+1}$. The last inequality always holds as $h_f^{n-s+1}$ is
a weighted average of the list of numbers $(1,\frac{1}{2},\frac{1}{3},...,\frac{1}{n-s+1})$ all of which are no less than $h_{\gamma}^{n-s+1}.$ \qed

~\

\noi The $\gamma$ core is based on the worst scenario for the members of the deviant coalition $S$: all outsiders in $N/S$ remain separate entities. Under our scenario, the singleton coalition structure is one only
of the structures that the members of $S$ take into account. Other, more favorable structures, occur with positive probability. Hence, under our framework, deviations from the grand coalition are "easier", which explains the relation between the two cores.

Let us conclude this section noticing that representations of the form (\ref{harmonic v})
hold for any probability distribution that a coalition assigns over the set of coalition structures.
Consider the distributions $g_{n,s}(j)$ and $z_{n,s}(j)$. Let
$(N,v_g)$ and $(N,v_z)$ denote the corresponding games and let $C_g$ and $C_z$ denote the cores of
the two games. We have the following

\vsp\noi {\bf{Corollary 1}} \emph{Consider two probability distributions $g_{n,s}(j)$ and $z_{n,s}(j)$ such that
$h_{g_{n,s}}^{n-s+1}> h_{z_{n,s}}^{n-s+1}$, for all $s$. If $C_g\neq\emptyset$ then $C_z\neq\emptyset$ as well.}

\vspace{0.2cm}\noi \textbf{Proof.} Since $h_{g_{n,s}}^{n-s+1}> h_{z_{n,s}}^{n-s+1}$ then $v_g(S)> v_z(S)$. Let $x\in C_g$. Then
$\sum\limits_{i\in S}x_i\geq v_g(S)>v_z(S)$ for any $S$. Hence $x\in C_z$ and $C_z\neq\emptyset$. \qed

~\

\noi Corollary 1 is useful in allowing for a operational comparison of the cores of two (or more) different games.
If we know that the core of one of the two games is non-empty, Corollary 1 gives us a (convenient) sufficient condition for the non-emptiness of the core of the other: we simply need to compare two numbers, i.e., the probabilistic harmonic numbers induced by the corresponding distributions.

\section{Conclusions}
This paper has incorporated elements of bounded rationality
into the study of cooperative oligopoly games with externalities. When a coalition
contemplates to not cooperate with the rest of the players it assumes that all possible coalition structures can form with positive probability. Whenever the number of firms is sufficiently large, then the core of the game is non-empty. Furthermore, it is a subset of the $\gamma$ core.

We assumed that all coalition structures occur according to the probability distribution
$f_{n,s}$. Other probability schemes can produce core non-emptiness as well. Consider distributions under which
the probabilities assigned to coalition structures with relatively many coalitions are higher compared to $f_{n,s}.$ Clearly, under these distributions, the core will be non-empty more often, i.e., for more values of $n$.
The reason is that these distributions penalize the structures that are more favorable
for a deviant coalition (i.e, structures with few coalitions) and give more weight to
less favorable structure (i.e., structures with many coalitions).

Finally, let us mention a few extensions of the current work.
The analysis of oligopolistic markets with more general demand and cost functions
and/or other modes of competition (e.g., product differentiation, price competition)
are natural future directions. Further, the application of
the current framework to other economic environments (e.g, environmental agreements, etc.) or to abstract cooperative games with externalities is of special interest.

\section*{References}

\noi 1. Aumann, R. (1959). Acceptable points in general cooperative n-person games,
Contributions to the theory of games IV, Annals of Mathematics
Studies vol. 40, Princeton University Press, Princeton.

\vspace{0.2cm}\noi 2. Bell E. T. (1934). Exponential Numbers.
American Mathematical Monthly, 41, 411-419.

\vspace{0.2cm}\noi 3. Bloch F. (1996). Sequential formation of
coalitions in games with externalities and fixed payoff division, Games and Economic Behavior, 14
90-123.

\vspace{0.2cm}\noi 4. Chander, P. and H. Tulkens (1997). A core of an economy with multilateral environmental
externalities, International Journal of Game Theory 26, 379-401.

\vspace{0.2cm}\noi 5. Currarini S. and M. Marini (1998). The core of games with Stackelberg leaders,
Working Paper, MPRA.

\vspace{0.2cm}\noi 6. Rajan R. (1989).
Endogenous coalition formation in cooperative oligopolies,
International Economic Review, 30, 4, 863-876.

\vspace{0.2cm}\noi 7. Ray D. and R. Vohra (1999). A theory of endogenous coalition structures,
Games and Economic Behavior, 26, 286-336.

\vspace{0.2cm}\noi 8. Sandholm, T., Larson, K., Andersson, M., Shehory, O., Tohm\'{e}, F. (1999). Coalition
structure generation with worst case guarantees. Artificial Intelligence
111(1-2), 209-238.

\vspace{0.2cm}\noi 9. Zhao, J. (1999). A $\beta$-core existence result and its application to oligopoly markets, Games
and Economic Behavior, 27, 153-168.

\section*{Appendix}

\begin{table}[h]
\begin{center}
\small
\begin{tabular}{|r|r|}
\hline
$s$&$\frac{v(s)}{(a-c)^2}$\\
\hline
1&0.0226\\
2&0.0252\\
3&0.0285\\
4&0.0326\\
5&0.0378\\
6&0.0446\\
7&0.0539\\
8&0.0672\\
9&0.0865\\
10&0.1111\\
11&0.25\\

\hline
\end{tabular}
\caption{values $v(s)$ of coalition $S:|S|=s$ in a game with $n=11$ players.}
\label{nonempty}
\end{center}
\end{table}

\begin{table}[t]
\begin{center}
\small

\begin{tabular}{|r|r|}
\hline
$n$&$\frac{v(\{i\})}{(a-c)^2}$\\
\hline
3&0.0865\\
4&0.0672\\
5&0.0539\\
6&0.0446\\
7&0.0378\\
8&0.0326\\
9&0.0285\\
10&0.0252\\

\hline
\end{tabular}
\caption{values $v(\{i\})$, $n \in  \{3,4,\cdots,10\}$}
\label{n=11_table}
\end{center}
\end{table}

\end{document}